\documentclass[a4paper,twoside]{article}

\usepackage{epsfig}
\usepackage{subcaption}
\usepackage{calc}
\usepackage{amssymb}
\usepackage{amstext}
\usepackage{amsmath}
\usepackage{amsthm}
\usepackage{multicol}
\usepackage{pslatex}
\usepackage{apalike}
\usepackage[bottom]{footmisc}
\usepackage{SCITEPRESS}     

\usepackage{graphicx}
\usepackage[dvipsnames]{xcolor}
\usepackage{algorithm}
\usepackage{algcompatible}
\usepackage{multirow}
\usepackage{todonotes}
\usepackage{adjustbox}
\usepackage{xcolor}

\newtheorem{definition}{Definition}[section]

\def\bs{\boldsymbol}

\begin{document}

\title{Geodesic Gramian Denoising Applied to the Images Contaminated With Noise Sampled From Diverse Probability Distributions}
\author{\authorname{Yonggi Park \sup{1}\orcidAuthor{0000-0003-3544-4413}, Kelum Gajamannage\sup{1}\orcidAuthor{0000-0001-9179-3787}, and Alexey Sadovski \sup{1}\orcidAuthor{0000-0001-7101-3632}}
\affiliation{\sup{1} Department of Mathematics \& Statistics, Texas A\&M University-Corpus Christi, 6300 Ocean Drive, Corpus Christi TX-78412, USA}
\email{yonggi.park@tamucc.edu, kelum.gajamannage@tamucc.edu, alexey.sadovski@tamucc.edu}
}

\keywords{Denoising, Probability Distributions, Gramian, Geodesics, Patch-base.}

\abstract{
As quotidian use of sophisticated cameras surges, people in modern society are more interested in capturing fine-quality images. However, the quality of the images might be inferior to people's expectations due to the noise contamination in the images. Thus, filtering out the noise while preserving vital image features is an essential requirement. Current existing denoising methods have their own assumptions on the probability distribution in which the contaminated noise is sampled for the method to attain its expected denoising performance. In this paper, we utilize our recent Gramian-based filtering scheme to remove noise sampled from five prominent probability distributions from selected images. This method preserves image smoothness by adopting patches partitioned from the image, rather than pixels, and retains vital image features by performing denoising on the manifold underlying the patch space rather than in the image domain. We validate its denoising performance, using three benchmark computer vision test images applied to two state-of-the-art denoising methods, namely BM3D and K-SVD.}

\onecolumn \maketitle \normalsize \setcounter{footnote}{0} \vfill

\section{\uppercase{Introduction}}\label{sec:intro}
\label{sec:intro}
The need for robust and accurate image denoising methods has emerged with the massive production of digital images because many images are often captured by imprecise devices or under poor environmental conditions \cite{buades2010image}. Image analysis tasks, such as classification and feature extraction, may be downgraded due to the existence of noise in images; thus, denoising is an inevitable step before performing image analysis. Since noise is considered to be the high-frequency component of a real-life noisy image, an exact separation of the noise from an image is implausible \cite{fan2019brief}. Thus, the probability distribution governing the noise contaminated image and the amount of noise corruption are concealed. 
The denoising schemes available in literature not only make nonviable assumptions about the noise distributions, such as Gaussian, but also require the user to provide the intensity of the image's noise corruption. Thus, the denoising schemes often lose the vital image information contained in the original image.  \cite{fan2019brief}.

Image processing literature states denoising images under contaminations can be sampled from diverse probability distributions. The noise types, Gaussian, salt and pepper, speckle, Poisson, and uniform, are frequently employed in such denoising tasks \cite{makandar2014comparative}. 

Gaussian noise is often caused by natural sources, such as thermal vibration of atoms and discrete nature of warm objects radiation; however, it is also present in electronic devices, such as amplifiers and detectors \cite{boyat2015review}. Salt and pepper noise presents as white and black pixels, respectively, on an image that occurs sparsely and randomly which is seen in data transmission \cite{hosseini2013fast}. Speckle noise is multiplicative and fluctuates nonidentically on each cell such that it is highly sensitive to small variations where this noise is seen in coherent imaging systems, such as laser and radar \cite{boyat2015review}. Poisson noise occurs when statistical information, identified by a sensor, does not have a sufficient amount of photons counting in optical devices, which is also known as shot photon noise \cite{hasinoff2014photon}. Uniform noise is generated in an amplitude quantization process, such as converting analog data into digital data \cite{widrow2008quantization}.

Image denoising methods can be categorized into two types: patch-based and pixels-based. The patch-based denoising method partitions the given noisy image into partially overlapping blocks, whereas the pixel-based image denoising method works purely on pixels \cite{ali2020novel}. Block matching 3D algorithm \cite{Dabov2007}, denoted as BM3D, is a patch-based denoising technique that groups the patches into 3D data arrays, and it applies a modified sparse representation of the patches in the frequency domain. The patch-based denoising method K-Singular Value Decomposition (K-SVD) \cite{Elad2006}, is a learning algorithm for creating a dictionary for sparse representations using a singular value decomposition approach. These two widely used state-of-the-art denoising methods are considered to be baselines of denoising techniques and are compared to the performance of new denoising methods.

We have recently shown that the eigenvectors of the geodesic Gramian matrix of image patches, denoted as GGD, \cite{gajamannage2020patch}, are capable of approximating the original image of a noised image with high fidelity. First, GGD partitions the noisy image into partially overlapping moving square patches with a known length, say $\rho$, such that each patch is centered at one unique pixel of the image. Each patch is a point in a $\rho^2$-dimensional ambient space where a hidden low-dimensional manifold, describing the prominent features of the noisy image, underlies \cite{gajamannage2015a}, \cite{gajamannage2015identifying}. Revealing this hidden manifold helps to better explain the structure of the patch set and then helps to identify the features in the image \cite{gajamannage2021bounded}. Revealing such a manifold is a nonlinear dimensionality reduction approach \cite{gajamannage2019nonlinear} where the extra dimensions mostly represent noise and minor image features. As GGD is focused on an eigenvalues analysis based on an underlying manifold of the patch space, this denoising scheme is considered to be a patch-based non-local dictionary learning method.

In this paper, we examine the denoising performance of our GGD method, and apply it to images contaminated with five diverse noise types sampled from unique probability distributions. These probability distributions are Gaussian, salt and pepper, speckle, Poisson, and uniform. To facilitate consistent comparison between diverse noise types, we use the relative norm of noise for each type instead of the pure noise contamination. For three relative noise levels, $30\%$, $40\%$, and $50\%$, GGD's denoising performance is compared to that of BM3D and K-SVD using two performance metrics: peak signal-to-noise ratio (PSNR) and structural similarity index (SSIM).

The paper is organized as follows: we provide both 1) information about noise types and associated probability distributions and 2) the formulation of the denoising method GGD in Sec.~\ref{sec:method}. Then, we analyze the performance of GGD with that of BM3D and K-SVD based on three benchmark computer vision test images in Sec.~\ref{sec:analysis}. Finally, we provide a discussion, along with the conclusions in Sec.~\ref{sec:conclusion}.

\section{\uppercase{Method}}\label{sec:method}
Here, we present both the five probability distributions associated with five noise types (Gaussian, salt and pepper, speckle, Poisson, and uniform), and the formulation of the GGD method.

\subsection{Probability distributions for noise samples}\label{sec:prob}
Consider that we are given a noise-free image of size $n\times n$, denoted as $\mathcal{I}_{n\times n}$. For each of the five noise types, we draw a noise sample of size $n\times n$, denoted as $\mathcal{N}_{n\times n}$, from the corresponding probability density function (PDFs) that we present in the sequel. While the noisy image, denoted as $\mathcal{U}_{n\times n}$, contaminated with speckle noise is generated by pointwise multiplicative rule
\begin{equation}
\mathcal{U}_{n\times n}=\mathcal{I}_{n\times n}\odot \mathcal{N}_{n\times n},
\end{equation}
where $\odot$ is the pointwise multiplication operator. Noisy image with Gaussian noise, salt and pepper noise, Poisson noise, and uniform noise is generated by additive rule
\begin{equation}
\mathcal{U}_{n\times n}=\mathcal{I}_{n\times n}+\mathcal{N}_{n\times n},
\end{equation}

The PDF for Gaussian noise is given by 
\begin{equation}
P_g(z,\mu,\gamma) = {\frac{1}{\sigma \sqrt{2 \pi}} { e^{ \frac{ -{(z-\mu)}^2 } { 2 \sigma^2 } } } }
\end{equation}
where $z$ is gray-level, $\mu$ is the mean of the distribution, $\sigma$ is the standard deviation of the distribution. The PDF of salt and pepper noise is given by
\begin{equation}
P_{sp}(z,d) =
\begin{cases}
d, $ for $ z=0 $ (pepper, i.e., black color)$,\\
d, $ for $ z=255 $ (salt, i.e., white color)$,\\
1-2d, $ otherwise$,
\end{cases},
\end{equation}
where $d$ is density that which is same the probability associated with each of the cases $z=0$ and $z=255$ \cite{chan2005salt}. The PDF of speckle noise is given by 
\begin{equation}
P_s(z,\alpha,\theta) = \frac{z^{\alpha-1}e^{-z/\theta}}{(\alpha-1)! \ \theta^{\alpha}}
\end{equation}
where $\alpha$ and $\theta$ are shape and scale parameters associated with Gamma distribution \cite{intajag2006speckle}. The PDF of Poisson noise is given by
\begin{equation}
P_p(z,\lambda) = \frac{{\lambda}^z e^{-\lambda}}{z!}
\end{equation}
where $\lambda$ is the expected number of occurrences per unit time \cite{hasinoff2014photon}. The PDF of the uniform noise is given by
\begin{equation}
P_u(z,a,b) =
\begin{cases}
\frac{1}{b-a}, $ if $ a \le z \le b\\
0, $ otherwise$
\end{cases},
\end{equation}
where $a$ and $b$ are the lower and upper bounds of the distribution.

Since our analysis is based on gray images, we transform the colored images into gray images by taking the average across the three color channels. The analysis here is focused on the images contaminated with five diverse noise types; thus, to make the noisy images consistent across their individual PDFs, we compute the relative error, denoted as $k$, by 
\begin{equation}\label{eqn:noise}
k=\frac{\|\mathcal{U}-\mathcal{I}\|_2}{\|\mathcal{I}\|_2} 100\%.
\end{equation}
For each noise type described here, the original noise-free image $\mathcal{I}_{100 \times 100}$ is imposed with diverse levels of noise intensities by changing the parameters of the corresponding PDF to create three noisy images with $k=30\%$, $40\%$, and $50\%$. Since $\mathcal{U}$'s represent images, the values of the pixels in them should be between 0--255 even after imposing the noise. Thus, we adjust $\mathcal{U}$ by replacing the values less than zero with zeros and the values more than 255 with 255's

\subsection{Geodesic Gramian denoising}
We denoise the images, that are imposed with five noise types, using geodesic Gramian denoising. First, GGD partitions the noisy images  $\mathcal{U}_{n\times n}$ into equal-sized square-shaped overlapping patches, denoted as $\bs{u}(\bs{x}_{ij})$'s; $i,j=1,\dots,n$, of odd length, denoted as $\rho$, centered at each pixel of the image. For simplicity, sometimes we write $\bs{u}(\bs{x}_k)$ for $\bs{u}(\bs{x}_{ij})$ where $k=n(i-1)+j$ and $1\le k \le n^2$. 

Second, we create a graph structure $G(V,E)$ on this dataset by defining the points, $\{\bs{u}(\bs{x}_k)\vert k=1,\dots,n^2\}$, as vertices, $V$. We define the edge set, $E$, by joining each pair of nearest neighbor points, say $\bs{u}(\bs{x}_{k})$ and $\bs{u}(\bs{x}_{k'})$, with an edge having the weight equal to the Euclidean distance, denoted as $d(k,k')$,
\begin{equation}\label{eqn:eudist}
d(k,k')=\|\bs{u}(\bs{x}_{k})-\bs{u}(\bs{x}_{k'})\|_2,
\end{equation} 
between them. GGD approximates geodesic distance between two patches in the patch-set as the shortest path distance between the corresponding two vertices in the graph $G(V, E)$ with the aid of Floyd’s algorithm \cite{floyd1962algorithm}. 

Third, GGD transforms the geodesic distance matrix, denoted as $\mathcal{D} \in\mathbb{R}^{n^2\times n^2}_{\ge0}$, into its Gramian matrix, denoted as $\mathcal{G}_{n^2 \times n^2}$, using
\begin{equation}\label{eqn:gram}
\mathcal{G}[i,j]=-\frac{1}{2}\big[\mathcal{D}[i,j]-\mu_i(\mathcal{D}) -\mu_j(\mathcal{D})+\mu_{ij}(\mathcal{D})\big],
\end{equation} 
where $\mu_i(\mathcal{D})$, $\mu_j(\mathcal{D})$, and $\mu_{ij}(\mathcal{D})$ are the means of the $i$-th row of the matrix $\mathcal{D}$, $j$-th column of that matrix, and the mean of the full matrix, respectively, \cite{lee2004nonlinear}. 

Fourth, the noisy patches, $\bs{u}_k$, are denoised using only $L$ prominent eigenvectors of the Gramian matrix because they represent essential features, but not the noise, of the image. We denote the denosed version of the patch $\bs{u}_k$ as $\tilde{\bs{u}}_k$ that GGD produces by
\begin{equation}\label{eqn:denoise}
\tilde{\bs{u}}(\bs{x}_k) = \sum^{L}_{l=1} \langle \bs{u}(\bs{x}_k), \bs{\nu}_l \rangle \bs{\nu}_l,
\end{equation}
where $\bs{\nu}_l$ represents $l$-th prominent eigenvector of the Gramian matrix where $l=1,\dots,L$, and $\langle \cdot, \cdot \rangle$ denotes the inner product.

Fifth, in order to construct the denosed image from the denoised patches, $\tilde{\bs{u}}$'s, GGD estimates the intensity of each pixel of the image using all the other pixels at the same location with respect to the location of the image but each exists in one of the $\rho^2$ overlapping patches. GGD combines all these estimations using a moving least square approximation given by using Shepard’s method \cite{Shepard1968} as
\begin{equation}\label{eqn:merge}
\tilde{\mathcal{U}}(\bs{x}_k)=\sum_{\bs{x}_t \in \mathcal{N}(\bs{x}_k)} \Gamma (\bs{x}_k,\bs{x}_t) [\tilde{\bs{u}}(\bs{x}_t)]_{t_n},
\end{equation}
where the neighborhood of the target pixel $\bs{x}_k$ is
\begin{equation}
\mathcal{N}(\bs{x}_k)=\{\bs{x}_{t} \ \vert \ \ \|\bs{x}_k-\bs{x}_{t}\|_{\infty}\le \rho/2\},
\end{equation}
\cite{Meyer2014} and the weight is given by
\begin{equation}\label{eqn:shepard_weight}
\Gamma(\bs{x}_k,\bs{x}_t)=\frac{e^{-\|\bs{x}_k-\bs{x}_t\|^2}}{\sum_{\bs{x}_{t'}\in \mathcal{N}(\bs{x}_k)} e^{-\|\bs{x}_k-\bs{x}_{t'}\|^2}}.
\end{equation}
The weighting term in Eqn.~\eqref{eqn:shepard_weight} weights close by pixels with more weight while the faraway pixels with less weight. Thus, according to Eqn.~\eqref{eqn:merge}, merging assures that the pixel $\bs{x}_k$ of the reconstructed image is highly influenced by the pixels at the same location of the nearby patches. The main steps of GGD are summarized in Algorithm~\ref{alg:algorithm}.

\begin{algorithm}[htp]
\caption{ 
Geodesic Gramian Denoising (GGD).\\
Inputs:  noisy image ($\mathcal{U}_{n\times n}$), patch length ($\rho$),\\
 nearest neighborhood size ($\delta$), and eigenvector threshold ($L$).\\
Outputs:  noise-reduced image ($\tilde{\mathcal{U}}_{n\times n}$).}
\begin{algorithmic}[1]
\STATE  Construct $n^2$ overlapping square-shaped patches each with the length $\rho$ from the noisy image $\mathcal{U}_{n\times n}$ and denote the patch set as $\{\bs{u}(\bs{x}_k)\vert k=1,\dots,n^2\}$.	
\STATE Produce the graph structure $G(V,E)$ from the patch set using the nearest neighbor search algorithm in \cite{agarwal1999geometric}. Use Floyd's algorithm in \cite{floyd1962algorithm}, to approximate the geodesic distances in the patch space and then produce the geodesic distance matrix $\mathcal{D}$.	
\STATE Construct the Gramian matrix $\mathcal{G}$ from the geodesic distance matrix $\mathcal{D}$ using Eqn.~\eqref{eqn:gram}.	
\STATE Compute the eigenvectors $\{\nu_l\vert l = 1,\dots L\}$ corresponding to the $L$ biggest eigenvalues of the Gramian matrix $\mathcal{G}$ and use Eqn.~\eqref{eqn:denoise} to produce noise-free patches $\{\tilde{\bs{u}}(\bs{x}_k)\vert k=1,\dots,n^2\}$.	
\STATE Merge noise-free patches using Eqns.~\eqref{eqn:merge} and \eqref{eqn:shepard_weight}, and generate the denoise image $\tilde{\mathcal{U}}_{n\times n}$.
\end{algorithmic}\label{alg:algorithm}
\end{algorithm}

\section{\uppercase{Performance Analysis}}\label{sec:analysis}
Here, we analyze the performance of GGD by both the visual perception and the two similarity metrics, namely, peak signal to noise ratio (PSNR), and structural similarity index measure (SSIM), and then compare GGD's performance with that of BM3D and K-SVD. We use three benchmark test images, namely, Indian, pepper, and fish \cite{sepas2014novel}, of size $100\times 100$.

\begin{figure*}[htp]
\includegraphics[width = 1.0\linewidth]{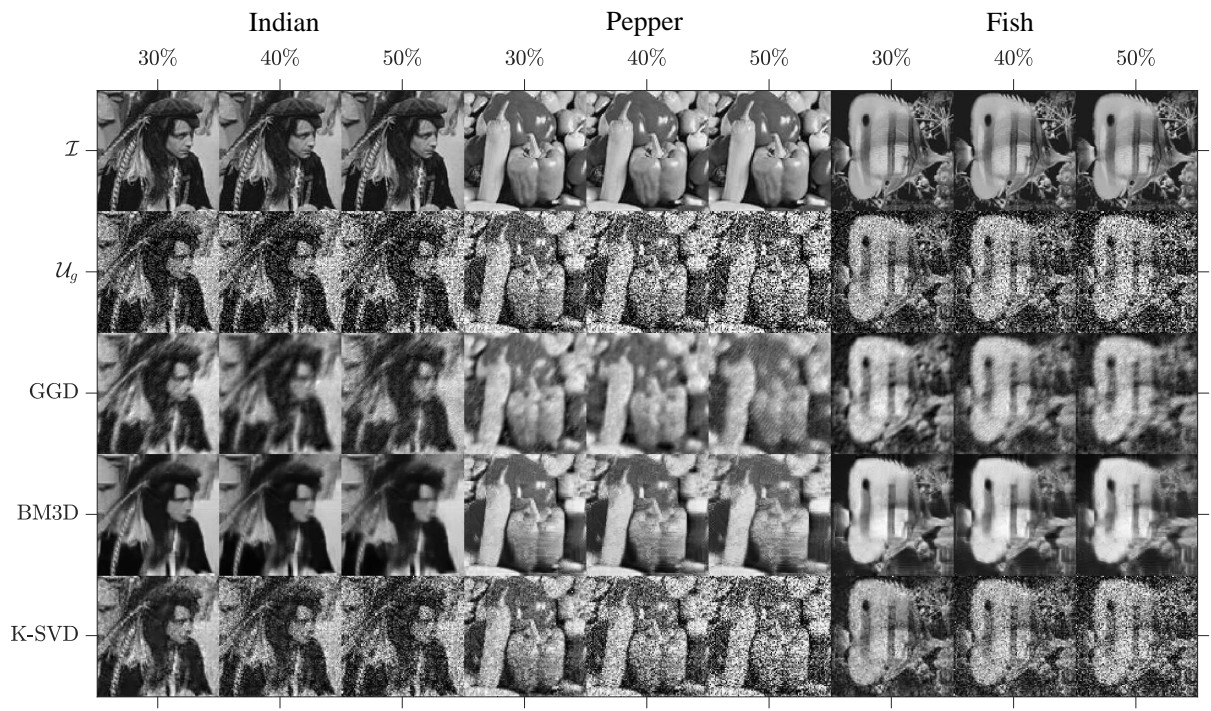}
\caption{Denoising images contaminated with the noise sampled from Gaussian PDF. Three test images, namely, Indian, Pepper, and Fish, denoted as $\mathcal{I}$, are contaminated with three relative noise levels, $30\%$, $40\%$, and $50\%$ , denoted as $\mathcal{U}_g$. The noisy images are denoised using GGD as well as two other state-of-the-art methods, BM3D and K-SVD.} 
\label{fig:psnrGauss}
\end{figure*}

\begin{figure*}[htp]
\includegraphics[width = 1.0\linewidth]{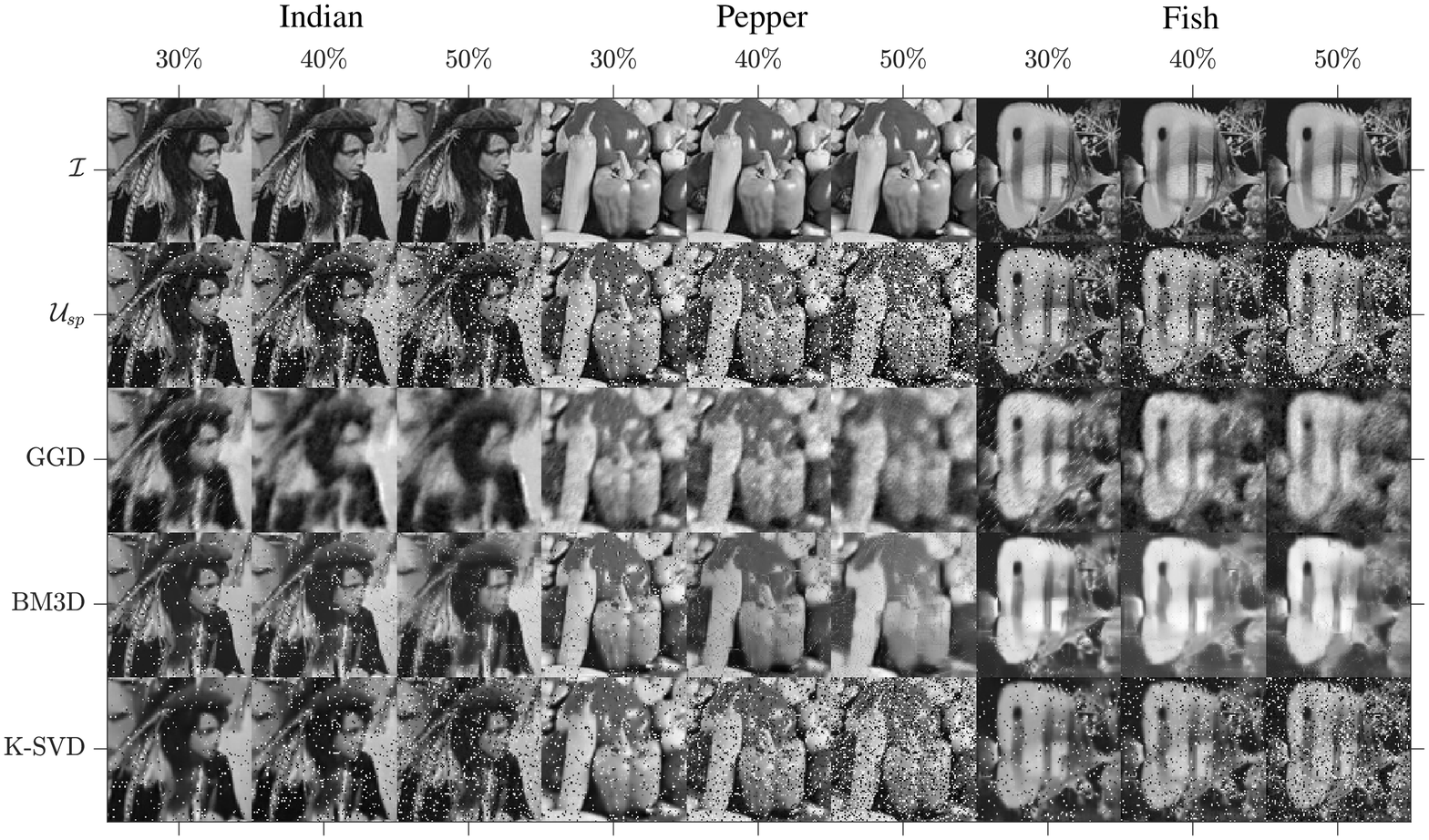}
\caption{Denoising images contaminated with the noise sampled from salt and pepper PDF. Three test images, namely, Indian, Pepper, and Fish, denoted as $\mathcal{I}$, are contaminated with three relative noise levels, $30\%$, $40\%$, and $50\%$ , denoted as $\mathcal{U}_{sp}$. The noisy images are denoised using GGD as well as two other state-of-the-art methods, BM3D and K-SVD.} 
\label{fig:psnrSP}
\end{figure*}

\begin{figure*}[htp]
\includegraphics[width = 1.0\linewidth]{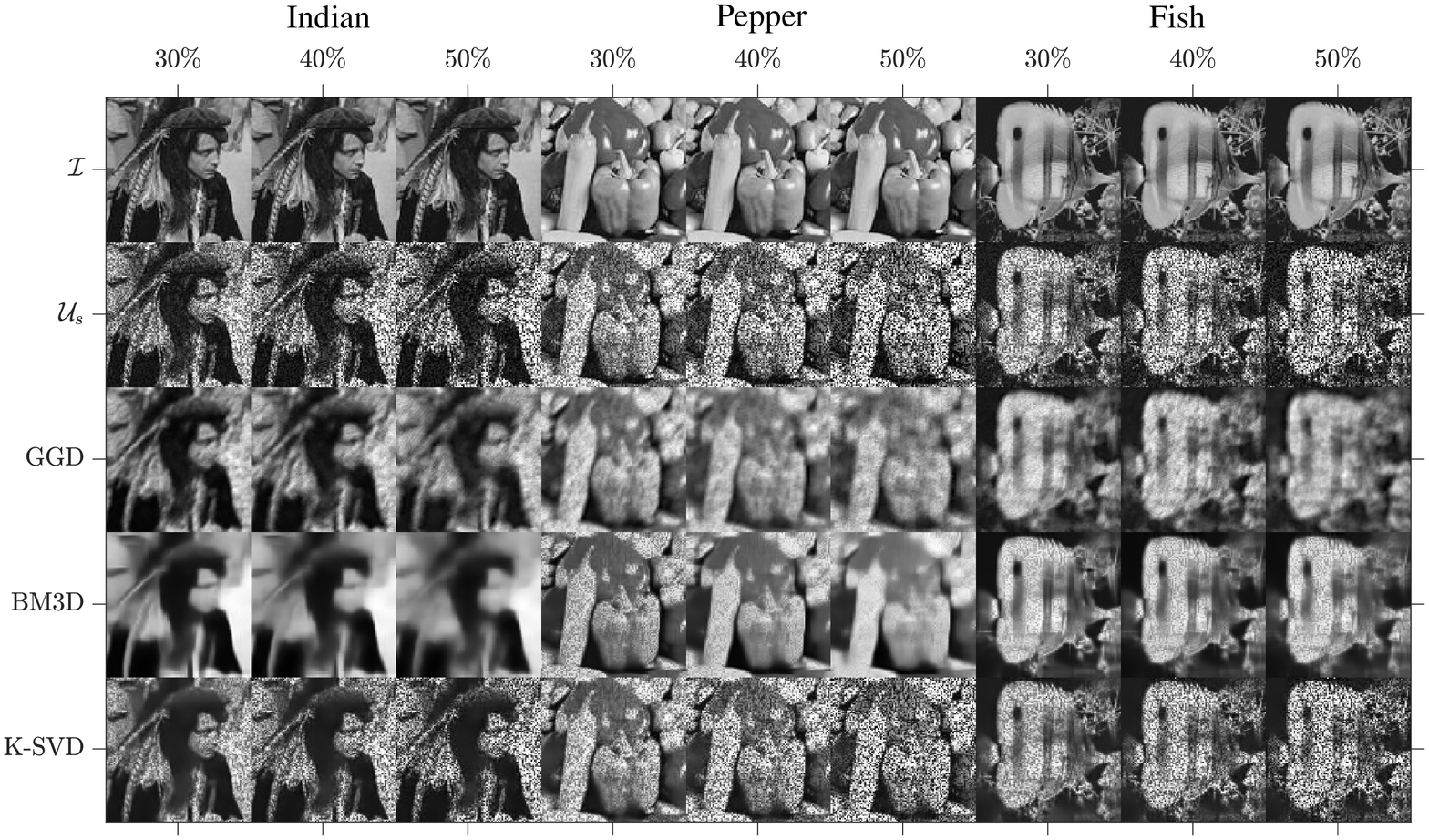}
\caption{Denoising images contaminated with the noise sampled from speckle PDF. Three test images, namely, Indian, Pepper, and Fish, denoted as $\mathcal{I}$, are contaminated with three relative noise levels, $30\%$, $40\%$, and $50\%$ , denoted as $\mathcal{U}_s$. The noisy images are denoised using GGD as well as two other state-of-the-art methods, BM3D and K-SVD.} 
\label{fig:psnrSpeckle}
\end{figure*}

\begin{figure*}[htp]
\includegraphics[width = 1.0\linewidth]{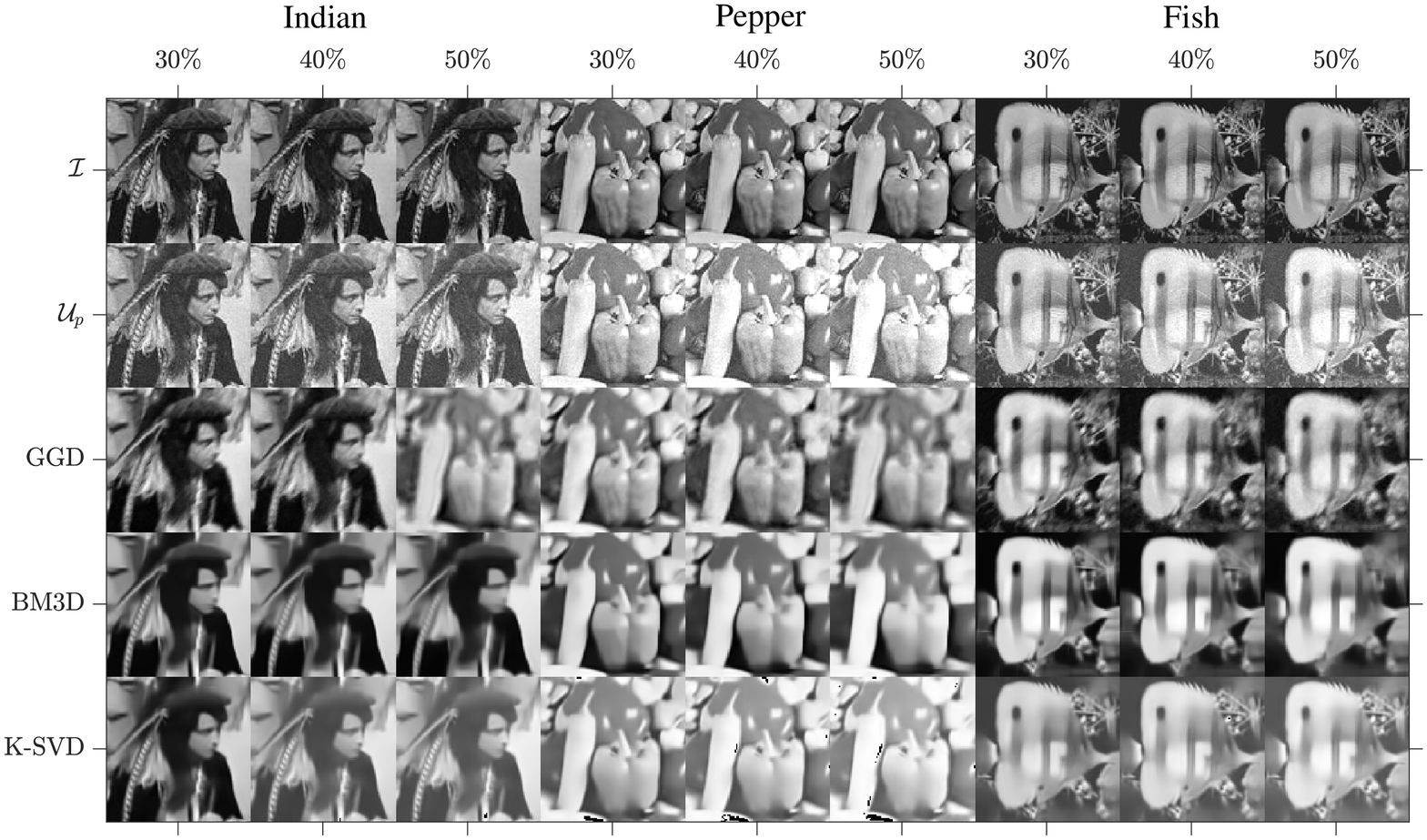}
\caption{Denoising images contaminated with the noise sampled from Poisson PDF. Three test images, namely, Indian, Pepper, and Fish, denoted as $\mathcal{I}$, are contaminated with three relative noise levels, $30\%$, $40\%$, and $50\%$ , denoted as $\mathcal{U}_p$. The noisy images are denoised using GGD as well as two other state-of-the-art methods, BM3D and K-SVD.} 
\label{fig:psnrPoisson}
\end{figure*}

\begin{figure*}[htp]
\includegraphics[width = 1.0\linewidth]{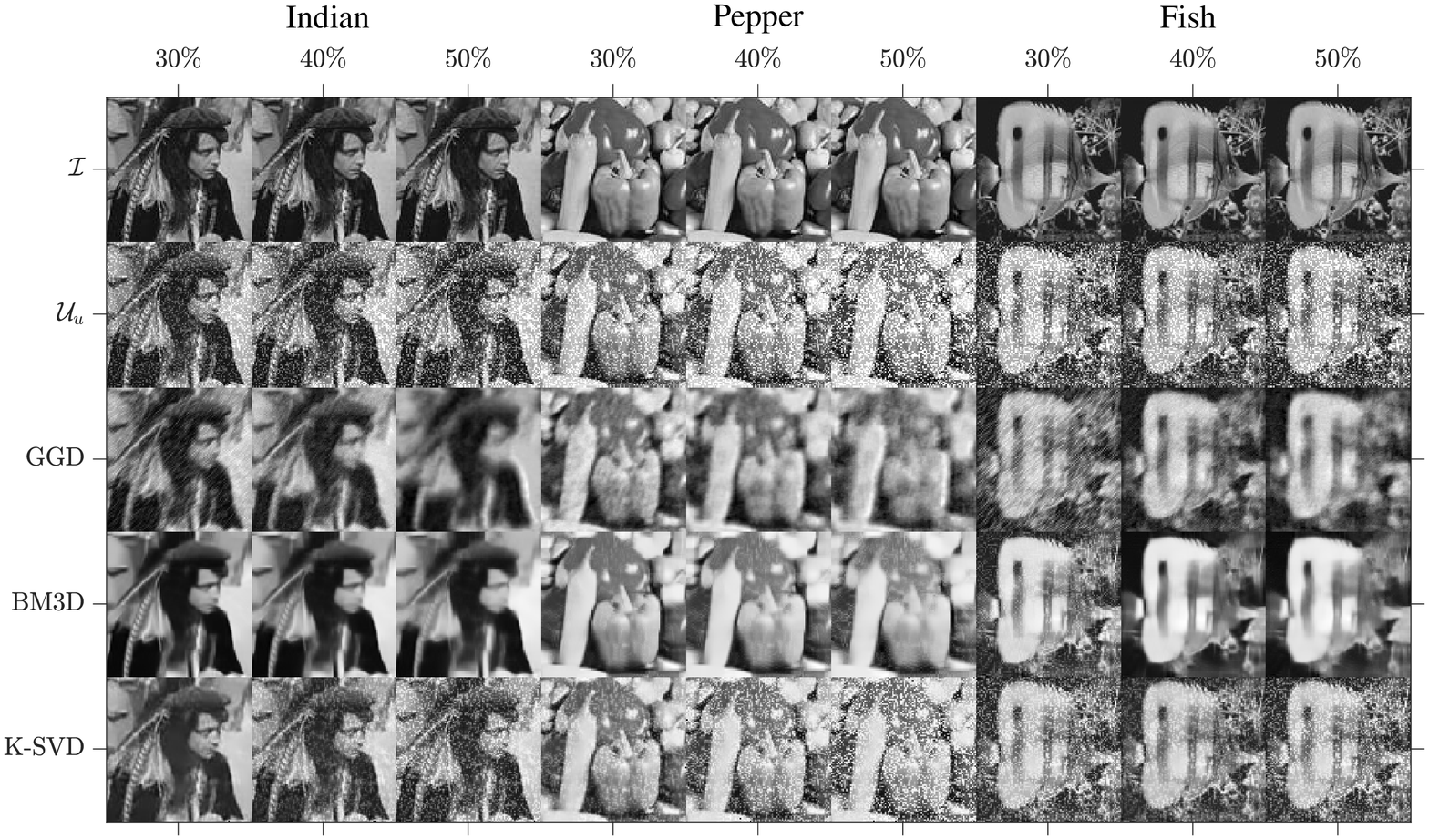}
\caption{Denoising images contaminated with the noise sampled from uniform PDF. Three test images, namely, Indian, Pepper, and Fish, denoted as $\mathcal{I}$, are contaminated with three relative noise levels, $30\%$, $40\%$, and $50\%$ , denoted as $\mathcal{U}_u$. The noisy images are denoised using GGD as well as two other state-of-the-art methods, BM3D and K-SVD.} 
\label{fig:psnrUniform}
\end{figure*}

After a superficial parameter tuning, for each noise type and each of the three images, the three parameters, neighborhood size ($\delta$), the patch size ($\rho$), and the eigenvector threshold ($L$), of GGD are set as ($\delta$, $\rho$, $L$) = (5,7,200), (10,9,100), and (15,11,50) for the noise levels $k$ = 30\%, 40\%, and 50\%, respectively. The baseline for such parameter selection is GGD performs better denoising with bigger neighborhood size, patch size, and eigenvector threshold for the images that are contaminated with bigger noise and it performs better denoising with smaller parameter values for the images  contaminated with smaller noise. We run GGD with noisy images, $\mathcal{U}$'s, and obtain the denoised versions of them, $\tilde{\mathcal{U}}$'s, see Figs.~\ref{fig:psnrGauss},~\ref{fig:psnrSP},~\ref{fig:psnrSpeckle},~\ref{fig:psnrPoisson}, and \ref{fig:psnrUniform}. We compute PSNR and SSIM between the noise-free image, $\mathcal{I}$, and the denoised image, $\hat{\mathcal{U}}$, using Def.~\ref{def:psnr} and Def.~\ref{def:ssim} that we show in Table~\ref{tab:psnrSSIM}. PSNR ranges between zero and $\infty$ such that zero ensures pure dissimilarity between $\mathcal{I}$ and $\mathcal{U}$ while $\infty$ ensures perfect similarity. SSIM ranges between -1 and 1 such that -1 indicates pure dissimilarity while 1 indicates perfect similarity. 

\begin{definition}\label{def:psnr}
Let, two-dimensional matrix $\mathcal{I}$ represents a reference image of size $n\times n$ and $\tilde{\mathcal{U}}$ represents any other image of interest. Peak Signal to Noise Ratio \cite{Hore2010}, abbreviated as PSNR, of the image $\tilde{\mathcal{U}}$ with respect to the reference image $\mathcal{I}$ is defined as  
\begin{equation}
PSNR(\mathcal{I},\tilde{\mathcal{U}}) = 20 \log_{10}\left(\frac{\max(\mathcal{I})}{RMSE(\mathcal{I},\tilde{\mathcal{U}})}\right).
\end{equation}
Here, $\max(\mathcal{I})$ represents the maximum possible pixel value of the image $\mathcal{I}$. Since the pixels in our images of interest are represented in 8-bit digits, $\max(\mathcal{I})$ is 255. PSNR ranges between zero and $\infty$ such that zero ensures pure dissimilarity between $\mathcal{I}$ and $\hat{\mathcal{U}}$ while $\infty$ ensures perfect similarity.
\end{definition}   

\begin{definition}\label{def:ssim}
Let, two-dimensional matrix $\mathcal{I}$ represents a reference image of size $n\times n$ and $\tilde{\mathcal{U}}$ represents an image of interest. Structural Similarity Index Measure \cite{Wang2004}, abbreviated as SSIM, of the image $\tilde{\mathcal{U}}$ with respect to the reference image $\mathcal{I}$ is defined as the product of luminance distortion ($I$), contrast distortion ($C$), and loss of correlation ($S$), such as
\begin{equation}
SSIM(\mathcal{I},\tilde{\mathcal{U}}) = I(\mathcal{I},\tilde{\mathcal{U}})C(\mathcal{I},\tilde{\mathcal{U}})S(\mathcal{I},\tilde{\mathcal{U}}),
\end{equation}
where
\begin{equation}
\begin{split}
I(\mathcal{I},\tilde{\mathcal{U}}) = \frac{2\mu_{\mathcal{I}}\mu_{\tilde{\mathcal{U}}}+c_1}{\mu^2_{\mathcal{I}}+\mu^2_{\tilde{\mathcal{U}}}+c_1},\\
C(\mathcal{I},\tilde{\mathcal{U}}) = \frac{2\sigma_{\mathcal{I}}\sigma_{\tilde{\mathcal{U}}}+c_2}{\sigma^2_{\mathcal{I}}+\sigma^2_{\tilde{\mathcal{U}}}+c_2},\\
S(\mathcal{I},\tilde{\mathcal{U}}) = \frac{\sigma_{\mathcal{I}\tilde{\mathcal{U}}}+c_3}{\sigma_{\mathcal{I}}\sigma_{\tilde{\mathcal{U}}}+c_3}.\\
\end{split}
\end{equation}
Here, $\mu_{\mathcal{I}}$ and $\mu_{\tilde{\mathcal{U}}}$ are means of $\mathcal{I}$ and $\tilde{\mathcal{U}}$, respectively; $\sigma_{\mathcal{I}}$ and $\sigma_{\tilde{\mathcal{U}}}$ are standard deviations of $\mathcal{I}$ and $\tilde{\mathcal{U}}$, respectively; and  $\sigma_{\mathcal{I}\tilde{\mathcal{U}}}$ is the covariance between $\mathcal{I}$ and $\tilde{\mathcal{U}}$. Moreover, $c_1$, $c_2$, and $c_3$ are very small positive constants to avoid the case of division by zero. SSIM ranges between -1 and 1 such that -1 indicates pure dissimilarity while 1 indicates perfect similarity. 
\end{definition}   

The parameters, patch size, sliding step size, the maximum number of similar blocks, radius for search block matching, step between two search locations, 2D thresholding, 3D thresholding, and threshold for the block-distance, of BM3D are set to their default values as 8, 4, 16, 39, 1, 2, 2.8, and 3000, respectively \cite{Dabov2007}. The parameter denoising strength of BM3D is the most critical parameter for denoising performance as it provides the noise contamination of the noisy images in terms of the standard deviation of the Gaussian noise. The standard deviations of the  Gaussian noise corresponding to the relative noise levels, 30\%, 40\%, and 50\% are 33, 45, and 58. Thus, for each of the three test images and for each noise type, we set the denoising strength of BM3D to 33, 45, and 58 for relative noise levels 30\%, 40\%, and 50\%, respectively. We run BM3D with all the noisy images and denoise them, see Figs.~\ref{fig:psnrGauss},~\ref{fig:psnrSP},~\ref{fig:psnrSpeckle},~\ref{fig:psnrPoisson}, and \ref{fig:psnrUniform}. We compute PSNR and SSIM between the noise-free image and the denoised image using Def.~\ref{def:psnr} and Def.~\ref{def:ssim} that we show in Table~\ref{tab:psnrSSIM}.

The parameters, block size, dictionary size, number of training iterations, noise gain, and number of nonzero coefficients of K-SVD are set to their default values 8, 256, 10, 1.15, and 2, respectively \cite{Elad2006}. The most vital parameter for K-SVD's denoising performance is the Lagrangian multiplier as it is associated with the standard deviation of the Gaussian noise contamination of the noisy image. Since the standard deviations of the Gaussian noise are 33, 45, and 58, corresponding to the relative noise levels, 30\%, 40\%, and 40\%, respectively, we set this parameter as 30/33, 30/45, and 30/58, for the relative noise levels, 30\%, 40\%, and 40\%, respectively \cite{Elad2006}. Now, We run K-SVD with all the noisy images and denoise them, see Figs.~\ref{fig:psnrGauss},~\ref{fig:psnrSP},~\ref{fig:psnrSpeckle},~\ref{fig:psnrPoisson}, and \ref{fig:psnrUniform}. We compute PSNR and SSIM between the noise-free image and the denoised image using Def.~\ref{def:psnr} and Def.~\ref{def:ssim} that we show in Table~\ref{tab:psnrSSIM}.

\begin{table*}[htp]
\centering
\caption{Denoising images, using three methods, namely, GGD, BM3D, and K-SVD, that are contaminated with the noise sampled from five PDF, Gaussian, salt and pepper, speckle, Poisson, and uniform. Here, three test images, namely, Indian, Pepper, and Fish, are contaminated with three relative noise levels, $30\%$, $40\%$, and $50\%$, and used for the experiment. For each of the noisy image, we compute PSNR, see the values without parenthesis, and SSIM, see the values within parenthesis, between the noise-free image and the denoised image. For each given noise level, test image, and noise type, we denote the best PSNR value over the three denoising methods by red color and the second best PSNR value by blue color. Thus, GGD, BM3D, and K-SVD attain the highest PSNRs for 62\%, 36\%, and 2\% for the times, respectively.}
 \begin{tabular}{|c |c|| c | c | c || c | c | c || c | c | c|} 
 \hline
 Noise                                   &   Method   & \multicolumn{3}{|c||}{Indian}     &  \multicolumn{3}{|c||}{Pepper}  &  \multicolumn{3}{|c|}{Fish}   \\
 \hline\hline
                                           &    $k$                      & $30\%$ & $40\%$ &    $50\%$ & $30\%$ & $40\%$ & $50\%$ & $30\%$ & $40\%$ &   $50\%$\\
\hline
\multirow{6}{*}{Gaussian} & \multirow{2}{*}{GGD} & 20.2                           & \textcolor{blue}{20.0}  & \textcolor{blue}{19.3}  & \textcolor{red}{20.5} & \textcolor{red}{19.5} & \textcolor{red}{18.4}    & 20.7            & \textcolor{red}{19.6} & \textcolor{red}{18.7}\\ 
                                      &                                 & (0.66)                               & (0.60)  & (0.54) & (0.64) & (0.59) & (0.45) &  (0.64) & (0.56)  & (0.47)\\ 
\cline{2-11}
                                      & \multirow{2}{*}{BM3D} & \textcolor{blue}{22.5}  &\textcolor{red}{20.7}  & \textcolor{red}{19.8} & 16.4                       & 14.7                        & \textcolor{blue}{14.7}  & \textcolor{red}{21.0}           & \textcolor{blue}{19.5}  & \textcolor{blue}{18.5}\\ 
                                      &                                 &(0.73)     & (0.64)     & (0.59)      & (0.41)       & (0.33) & (0.30)     & (0.74)      & (0.66)       & (0.58)\\ 
\cline{2-11}
                                      & \multirow{2}{*}{K-SVD} &\textcolor{red}{22.6}  & 18.4                               & 15.5                           & \textcolor{blue}{19.1} & \textcolor{blue}{15.0} & 12.5                          & \textcolor{blue}{20.9} &17.0                    & 14.2\\ 
                                      &                              & (0.70)                               &(0.48)     & (0.34)      & (0.53)      & (0.35) & (0.24)     & (0.64)      & (0.43)       & (0.30)\\ 
 \hline\hline
\multirow{6}{*}{} & \multirow{2}{*}{GGD}  & 18.7                            &\textcolor{red}{18.3}   &\textcolor{red}{18.6}  &\textcolor{red}{20.5}  &\textcolor{blue}{19.7}  &\textcolor{red}{18.5} &\textcolor{red}{19.3} & \textcolor{red}{19.4} & \textcolor{red}{18.5}\\ 
    			             &                                &  (0.57)    &(0.54)      &(0.51)       &(0.66)       &(0.59)  &(0.47) &   (0.58)   &(0.54) &(0.45) \\ 
\cline{2-11}
     Salt and                     & \multirow{2}{*}{BM3D} & \textcolor{red}{19.5}  & \textcolor{blue}{18.0}  & \textcolor{blue}{18.5} & \textcolor{blue}{20.2} & \textcolor{red}{20.0} &\textcolor{blue}{18.4}  & \textcolor{blue}{18.3} & \textcolor{blue}{15.6} & \textcolor{blue}{16.7}\\ 
         pepper                   &                                & (0.61)     & (0.55)     & (0.54)     & (0.63)       & (0.61) & (0.56)    & (0.69)      & (0.57)       & (0.54)\\ 
\cline{2-11}
                                      & \multirow{2}{*}{K-SVD} & \textcolor{blue}{19.4}  & 16.3                        & 15.2                          & 17.3                         & 15.0                         & 12.7                            & 18.0                     & \textcolor{blue}{15.8}                         & 14.0\\ 
                                      &                                & (0.55)     & (0.41)     & (0.36)      & (0.48)       & (0.37) & (0.25)     & (0.51)      & (0.39)       & (0.33)\\ 
 \hline\hline
\multirow{6}{*}{Speckle} & \multirow{2}{*}{GGD}  &\textcolor{red}{21.3}  &\textcolor{red}{20.2}   &\textcolor{red}{19.2}  &\textcolor{red}{21.3} & \textcolor{blue}{19.6} & \textcolor{red}{18.1}  &\textcolor{blue}{20.4} &\textcolor{blue}{19.6}  &\textcolor{blue}{18.7} \\ 
                                      &                                & (0.66) & (0.61)  &(0.52)  &(0.66)  &(0.58)  &(0.48)   &(0.63) &(0.57)  &(0.50)\\ 
\cline{2-11}
                                      & \multirow{2}{*}{BM3D} &17.7                         & 17.2                          & \textcolor{blue}{16.6} & \textcolor{blue}{20.2} & \textcolor{red}{20.5} & \textcolor{blue}{17.9}  &\textcolor{red}{21.2} & \textcolor{red}{19.7} & \textcolor{red}{19.0}\\ 
                                      &                                 & (0.60)      & (0.52)     & (0.47)      & (0.64)       & (0.62) & (0.57)    & (0.66)      & (0.57)       & (0.52)\\ 
\cline{2-11}
                                      & \multirow{2}{*}{K-SVD} &\textcolor{blue}{21.0}   &\textcolor{blue}{17.5}  & 15.1                       & 18.6                         & 15.0                          & 12.4                          & 19.5                           & 16.1 & 13.8\\ 
                                      &                               &(0.67)      & (0.59)     & (0.50)     & (0.57)       & (0.42) & (0.30)    & (0.64)      & (0.53)       & (0.44)\\ 
 \hline\hline
\multirow{6}{*}{Poisson} & \multirow{2}{*}{GGD}  & \textcolor{blue}{19.7}  & \textcolor{blue}{19.3}  & \textcolor{blue}{18.9} & \textcolor{blue}{19.5} & \textcolor{blue}{17.7} & \textcolor{red}{17.2}  &\textcolor{red}{21.6} & \textcolor{red}{21.0} & \textcolor{red}{20.4}\\ 
                                      &                                & (0.78)     & (0.70)    & (0.62)      & (0.80)       & (0.73) & (0.64)     & (0.79)      & (0.77)       & (0.75)\\ 
\cline{2-11}
                                      & \multirow{2}{*}{BM3D} & \textcolor{red}{21.6}  & \textcolor{red}{20.0}  & \textcolor{red}{19.3} & \textcolor{red}{20.4} & \textcolor{red}{18.5}                & \textcolor{blue}{15.3}  & 19.8              & \textcolor{blue}{19.1} & 17.9\\ 
                                      &                                 &(0.75)     & (0.71)     & (0.67)     & (0.79)       & (0.74) & (0.66)     & (0.73)      & (0.67)       & (0.61)\\ 
\cline{2-11}
                                      & \multirow{2}{*}{K-SVD} & 17.2                        & 15.0                            & 13.3                          & 15.0                         & 12.6                                         &  10.8   & \textcolor{blue}{21.3}                  & 15.0                          & \textcolor{blue}{19.8}\\ 
                                      &                                 & (0.68)     & (0.65)     & (0.62)      & (0.73)       & (0.67) &   (0.61)   & (0.76)      & (0.66)       & (0.76)\\ 
 \hline\hline
\multirow{6}{*}{Uniform} & \multirow{2}{*}{GGD}  & \textcolor{blue}{20.9}                      & \textcolor{blue}{18.5}  & \textcolor{red}{18.6} & \textcolor{red}{20.0} & \textcolor{red}{18.5} & \textcolor{red}{17.1}  & \textcolor{red}{17.9} &\textcolor{red}{19.7} & \textcolor{red}{18.4}\\ 
                                      &                                & (0.71)      & (0.64)     & (0.57)      & (0.70)       & (0.63) & (0.50)     & (0.56)      & (0.60)       & (0.55)\\ 
\cline{2-11}
                                      & \multirow{2}{*}{BM3D}&\textcolor{red}{21.1}  & \textcolor{red}{18.8}  & \textcolor{blue}{17.2} & \textcolor{blue}{19.4} & \textcolor{blue}{16.9} & \textcolor{blue}{15.2}                   &\textcolor{blue}{16.5}           & \textcolor{blue}{18.1}                         &\textcolor{blue}{17.1}\\ 
                                      &                                & (0.76)      & (0.71)     & (0.66)      & (0.77)       & (0.69) & (0.58)     & (0.61)      & (0.69)       & (0.63)\\ 
\cline{2-11}
                                      & \multirow{2}{*}{K-SVD} & 20.5  & 17.9                       & 15.3                           & 18.1                         & 14.9                             &   12.34                                                               & 15.2                        & 17.6                                 & 14.9\\ 
                                      &                               & (0.78)      & (0.63)     & (0.47)      & (0.69)       & (0.50) &  (0.37)    & (0.46)      & (0.61)       & (0.46)\\ 
 \hline
 \end{tabular}
\label{tab:psnrSSIM}
\end{table*}

Here, we vary three relative noise levels, $k=$ 30\%, 40\%, and 50\%, and five noise types, Gaussian, salt and pepper, speckle, Poisson, and uniform, for three test images, Indian, pepper, and fish, there are 45 different combinations of experiments that we executed using each of GGD, BM3D, and K-SVD. Figs.~\ref{fig:psnrGauss},~\ref{fig:psnrSP},~\ref{fig:psnrSpeckle},~\ref{fig:psnrPoisson}, and \ref{fig:psnrUniform} visualize GGD's denoising performance compared to that of BM3D and K-SVD. To quantitatively assess the performance, we computed PSNR and SSIM for all the denoised images as we show in Table~\ref{tab:psnrSSIM}. This table states that GGD, BM3D, and K-SVD attain the highest PSNR for 62\%, 36\%, and 2\% for the times. This comparison justifies that the order of performance from the best to worst is GGD, BM3D, and K-SVD. We also see that both PSNR and SSIM decrease as $k$ increases irrespective of the denoising method since more corruption reduces the denoising performance.

\section{\uppercase{Conclusion}}\label{sec:conclusion}
Eliminating the noise contaminated in real-life images is an arduous task due to multiple reasons such as unawareness of the intensity of the noise, noise is sampled from diverse PDFs, noise is not additive, etc. Our denoising method based on dictionary learning of geodesics' Gramian that is computed over the image patch space, denoted as GDD, is free from the assumptions about both the PDF of the contaminated noise or the constraints of noise integration. Moreover, GGD doesn't require parameter inputs that are associated with the intensity of the noise contamination such as denoising strength in BM3D and Lagrangian multiplier in K-SVD. In this research, we utilized five different PDFs, namely, Gaussian, salt and pepper, speckle, Poisson, and uniform, each generating a unique noise that can present in real-world images. In order to strengthen our analysis, we used two other state-of-the-art methods, BM3D and K-SVD, to denoise three test images, Indian, pepper, and fish, imposed with three relative noise levels, 30\%, 40\%, and 50\%. The denoised images are compared using visual perception and two similarity metrics, namely, PSNR and SSIM. 

We chose the parameters, neighborhood size, the patch size, and the eigenvector threshold of GGD with a superficial parameter tuning. We set denoising strength parameter in BM3D as the standard deviation, denotes as $\sigma$, of the Gaussian noise related to the corresponding relative noise \cite{Dabov2007}, and set the Lagrangian multiplier parameter in K-SVD to $30/\sigma$ \cite{Elad2006}. This parameter adoption procedure makes all the three denoising methods free from selection bias. We executed 45 different combinations of experiments where GGD, BM3D, and K-SVD attained the highest PSNR for 62\%, 36\%, and 2\% for the times. Thus, we can conclude that the order of performance from the best to worst is GGD, BM3D, and K-SVD.

GGD computes eigenvectors of a Gramian matrix of size $n^2 \times n^2$ for an image of $n\times n$ pixels where the computational complexity of the eigenvalue decomposition of this matrix is as big as $n^6$ \cite{Lee2009}. GGD always performs better denoising with small eigenvector thresholds; thus, the computation of only the required fewer number of eigenvectors of this Gramian matrix is worthwhile 
to overcome this big computational complexity issue. In the future, we will improve GGD by incorporating eigenvector estimation strategies that are available in the literature of \emph{Bigdata}. For that, we are planning to replace the regular eigenvalue decomposition routing with multiple eigenvalue approximation strategies such as, 1) inverse-free preconditioned \emph{Krylov subspace method} called \emph{Lanczos algorithm} \cite{Liang2014,Saad2011}; 2) \emph{random sampling} method that trains a neural network of rows of the matrix \cite{Kobayashi2001}; 3) \emph{Monte Carlo} approach that iteratively makes approximations \cite{Liang2014,Friedland2006}. This future work will reduce the computational complexity of GGD significantly so that the method can be brought into a stage where it can be integrated into real-time image denoising applications.

In this paper, we denoised images, contaminated with five noise types each sampled from a unique PDF, using our recent denoising method that is based on dictionary learning on geodesics' Gramian matrix computed over the image patch space. Our method is based on a manifold learning approach that does not assume any specific PDF for the noise sample or it doesn't require any parameter input related to the intensity of the noise. The denoising performance computed using the similarity matrices PSNR and SSIM justifies the validity of our method over two state-of-the-art methods, BM3D and K-SVD.  

\bibliographystyle{apalike}
{\small

}

\end{document}